\documentclass[%
reprint,
superscriptaddress,
 aps,
 pra,
longbibliography,
dvipsnames,
twocolumns,
xcolor=table
]{revtex4-1}
\usepackage[main=english]{babel}

\usepackage{dcolumn}
\usepackage{bm}
\usepackage{siunitx}
\usepackage{dsfont}
\usepackage{caption}
\usepackage{braket}
\usepackage{amsthm}
\usepackage{mathtools}
\usepackage{physics}
\usepackage{comment}
\usepackage{multirow}
\usepackage{lipsum}
\usepackage{algpseudocode}
\usepackage{amsthm}
\usepackage{algorithm}

\usepackage{tikz}
\usetikzlibrary{quantikz}

\usepackage{mathtools}
\usepackage{ragged2e}
\usepackage{subcaption}

\usepackage[utf8]{inputenc}
\usepackage[toc,page]{appendix}

\usepackage{graphicx}
\usepackage{hyperref}
\hypersetup{colorlinks}
\usepackage{soul}

\usepackage{bbold}
\usepackage{amssymb}
\usepackage{float}
\usepackage{hyphenat}

\hypersetup{colorlinks=true,citecolor=blue,linkcolor=blue,filecolor=blue,urlcolor=blue,breaklinks=true}

\begin{document}

\newcommand{\orcidicon}[1]{\href{https://orcid.org/#1}{\includegraphics[height=\fontcharht\font`\B]{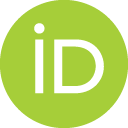}}}

\author{Francesco~Pio~Barone\orcidicon{0000-0003-2689-8499}}
\affiliation{European Organization for Nuclear Research (CERN), Geneva 1211, Switzerland}
\affiliation{University of Padua, 35122 Padua, Italy}

\author{Oriel~Kiss\orcidicon{0000-0001-7461-3342}}
\affiliation{European Organization for Nuclear Research (CERN), Geneva 1211, Switzerland}
\affiliation{Department of Nuclear and Particle Physics, University of Geneva, Geneva 1211, Switzerland}

\author{Michele~Grossi\orcidicon{0000-0003-1718-1314}}
\email{michele.grossi@cern.ch}
\affiliation{European Organization for Nuclear Research (CERN), Geneva 1211, Switzerland}

\author{Sofia~Vallecorsa\orcidicon{0000-0002-7003-5765}}
\affiliation{European Organization for Nuclear Research (CERN), Geneva 1211, Switzerland}

\author{Antonio~Mandarino\orcidicon{0000-0003-3745-5204}}
\affiliation{International Centre for Theory of Quantum Technologies, University of Gdańsk, Jana Bażyńskiego 1A, 80-309 Gdańsk, Poland}

\hypersetup{colorlinks=true,citecolor=magenta,linkcolor=magenta,filecolor=magenta,urlcolor=magenta,breaklinks=true}

\title{Counterdiabatic optimized driving in quantum phase sensitive models}

\date{\today}

\begin{abstract} 
State preparation plays a pivotal role in numerous quantum algorithms, including quantum phase estimation. This paper extends and benchmarks counterdiabatic driving protocols across three one-dimensional spin systems characterized by phase transitions: the axial next-nearest neighbor Ising (ANNNI), XXZ, and Haldane-Shastry (HS) models. We perform quantum optimal control protocols by optimizing the energy cost function, which can always be evaluated as opposed to the fidelity one requiring the exact state. Moreover, we incorporate Bayesian optimization within a code package for computing various adiabatic gauge potentials. This protocol consistently surpasses standard annealing schedules, often achieving performance improvements of several orders of magnitude. Notably, the ANNNI model stands out as a notable example, where fidelities exceeding 0.5 are attainable in most cases. Furthermore, the optimized paths exhibits promising generalization capabilities to higher-dimensional systems, allowing for the extension of parameters from smaller models. This opens up possibilities for applying the protocol to higher-dimensional systems. However, our investigations reveal limitations in the case of the XXZ and HS models, particularly when transitioning away from the ferromagnetic phase. This suggests that finding  optimal diabatic gauge potentials for specific systems remains an important research direction. 

\end{abstract}

\maketitle 
\section{Introduction}
The control of time-dependent dynamics in quantum systems is a crucial subroutine in many applications. In these procedures, undesired state transitions, e.g. between the instantaneous eigenstates of the external driver Hamiltonian, pose significant obstacles to maintaining the fidelity of the quantum state. This underscores the reliance on adiabatic dynamics in many control protocols, where the system faithfully follows the instantaneous eigenstates, inherently preventing unwanted transitions. Adiabatic processes, in their ideal form, are entirely reversible, rendering them theoretically robust. However, they usually require to go very slow, which can be an obstacle for most applications \cite{acin2017european}. 

One of the most foreseen applications of adiabatic quantum computing is state preparation (ASP), which is an input to many quantum algorithms such as quantum phase estimation (QPE) \cite{QPE,QPE-Lloyd} or variant based on time series analysis \cite{somma,LT,wang2022quantum,QETU}. Indeed, the quality of the initial state is a crucial ingredient and their preparation remains an important research question. It has to be noted that, in such cases, useful information could be retrieved even when the fidelity between the target and prepared state is above some threshold $\eta <1$, motivating the benchmarking and development of approximate ASP procedure.

The literature on this topic includes a variety of approaches aiming to speed-up the system dynamics and improve its accuracy. They can be broadly classified into two main classes. The first one includes all the techniques based on quantum optimal control (QOC) \cite{optimal_control_Glaser,optimal_control_introduction}, aiming at implementing optimal driving control, such as finding the optimal path avoiding any gap closure \cite{ASP_interpolation}, e.g. using machine learning \cite{optimal_control_ML}. The second one is rooted in shortcuts to adiabaticity (STA) protocols \cite{STA,STA_review}, whose strategy is the suppression of diabatic transitions between the instantaneous eigenstates of the dynamical Hamiltonian. One particularly successful technique in this context is given by counteradiabatic driving (CD) \cite{CD1,CD2,CD_Berry}, which reduces losses occurring when the system undergoes fast deformations, far from the adiabatic limit, by analytical compensation. While being an exact method, CD requires the solution of the full Schrödinger equation, which is not known in general. However, this condition can be relaxed by instead using an ansatz, which mitigates the diabatic losses, instead of canceling them completely. This includes variational approaches for the construction of the CD term \cite{variational_AGP_Campbell,variational_AGP_Saberi}, using neural networks \cite{CD_ML1,CD_ML2,CD_PIML}, or achieved by local counterdiabatic driving (LCD) \cite{LCD_Dries,LCD_Gjonbalaj,LCD_Kolodrubetz}, which are built using local interactions. It is important to note that these approximate methods are not guaranteed to succeed \cite{failure_ASP}, making it important to find good CD terms for each physical system. 

In this work, we will make use of the Counterdiabatic Optimal Local Driving (COLD) \cite{COLD} approach, which blends together QOC and LCD. 
One of our aims is to provide the physics community with a ready-to-use toolkit, that can serve the needs of several research areas 
dealing with complex statistical physics and for which quantum simulations constitute a valuable resource. 

The COLD protocol is benchmarked on three non-trivial spin models that serve as prototypes for application in condensed matter. 
The first one is the axial next-nearest neighbor Ising (ANNNI), a non-integrable extension of the Ising model, 
 describing spatial modulated magnetic patterns in crystals and alloys. Subsequently, application of COLD are studied on two different deformations of the Heisenberg model, that nevertheless being integrable, show rich and diversified phase diagrams: the XXZ model, and the Haldane-Shastry (HS) closed chain. In the former, the $SU(2)$ symmetry is broken down to the $U(1)$ via a uni-axial anisotropy in the spin-spin interaction,  while the latter has an interaction that is inversely proportional to the square distance among particles.
All the models are notoriously problematic for adiabatic computing, due to gapless transitions. 
Besides benchmarking, a few side enhancements to COLD are provided in order to produce a framework that is applicable to general spin systems. 
For instance, we provide a python package \cite{colder} to swiftly compute the LCD, propose an optimization strategy based on Bayesian optimization, and find that the optimized paths can generalize to larger models in the case of the ANNNI model. We start by describing the tools, such as QOC in Sec.~\ref{QOC}, CD in Sec.~\ref{CD}, and COLD in Sec.~\ref{COLD}. The results on different models exhibiting quantum phase transitions are then shown and discussed in Sec.~\ref{results}.

\section{Methods}


The Adiabatic State Preparation (ASP) and Quantum Annealing (QA) protocols are designed to find the ground state of a given Hamiltonian. They usually start in the ground state of an easier system and slowly transition towards the Hamiltonian of interest. In this context, an optimal time-dependent control is a fundamental requirement. 

Both protocols can be described within a time-evolution paradigm. 
Let $H_0(t)$ be the time-dependent Hamiltonian acting on a quantum system. Suppose that the evolution starts at some initial time $t=t_i=0$, without loss of generality, in the ground-state $\ket{\psi_i}$ of the Hamiltonian $H_0(t_i)$.
Similarly, let us call the ground-state of the instantaneous Hamiltonian at a time $t=t_f > t_i$ , i.e. the ``target state'', $\ket{\psi_T}$.
The total evolution time is then given by $\tau\equiv t_f-t_i$.

Even though the setup is quite general and allows to solve a variety of optimization problems, it is subject to limitations. Indeed, both protocols rely on a crucial assumption, which is that the system follows the instantaneous ground state of $H_0(t)$ during its evolution, eventually reaching the target state with null infidelity. The backbone of this assumption is the ``adiabatic theorem'', which characterize the constraints under which the assumption holds true.

In general, the main requirements are that the transformation is infinitesimally slow ($\tau\rightarrow\infty$) and the instantaneous energy gap between the ground-state and the excited states is non-vanishing.

For example, Refs \cite{vanvreumingen2023adiabatic, Jansen_2007} report an instance of the adiabatic theorem that is specific to ASP. 
Let us call $\ket{\phi(s)}$ the state evolved on the quantum computer at time $s$, and $\ket{\psi(s)}$ the instantaneous ground-state of the Hamiltonian $H(s)$ at the same time. If $\ket{\psi(s)}$ is separated by a non-zero gap $\Delta(s)>0$ from the excited spectrum, then to guarantee the convergence towards the target state $|\langle\psi(s)|\phi(s)\rangle|\ge 1-\delta$, $\forall s\in[0,1]$, we require that
\begin{equation*}
    \tau \ge \frac{1}{\delta}
    \left( \int_0^s \left[
    \frac{||\partial_s^2H(\sigma)||}{\Delta^2(s)}
    + 7 \frac{||\partial_sH(\sigma)||^2}{\Delta^3(s)}
    \right] d\sigma + B \right)
    \;\;.
\end{equation*}

A straightforward consequence is that ASP (or QA) protocols require a long evolution time to achieve higher fidelities.
Even so, in the presence of a gap closure, the assumptions do not hold anymore and the system evolves naturally towards excited states. For instance, a gap closure occurs spontaneously in systems that exhibit phase transitions.

In the context of state preparation and annealing, the most relevant accuracy metric is the final fidelity of the prepared state. In the following sections, two groups of methods that aim to achieve that goal are introduced, which are fundamentally different but still compatible. 


\subsection{Quantum Optimal Control}
\label{QOC}
In Quantum Optimal Control \cite{optimal_control_Glaser,optimal_control_introduction}, the dynamics of the system is manipulated to optimize a given metric, such as the fidelity or the energy.
In the general time-evolution setup previously introduced, the system is prepared in the initial state $\ket{\psi_i}$ and evolves in time towards the target state $\ket{\psi_T}$.
A QOC problem is thus framed as the optimization of the Schrödinger equation
$$\dot\psi=f(t,\boldsymbol{\beta})\;\;,$$
where $\psi$ is the quantum wave function and $\boldsymbol{\beta}$ is the set of tunable control parameters. 

The choice of the metric plays an important role as it defines the optimization landscape. The most popular cost function in the context of ASP is the infidelity with respect to the target state, namely:
\begin{equation}
\label{eq:loss_fidelity}
    \mathcal{C}_f(\boldsymbol{\beta}) \equiv 1 - |\langle\psi_T|\psi_f(\boldsymbol{\beta})\rangle|^2 \in [0.1]
    \;.
\end{equation}

Rephrasing the dynamics in terms of Hamiltonians, the system undergoes an evolution controlled by the time-dependent Hamiltonian
\begin{equation}
    \label{eq:QOCHamiltonian}
    H_{\text{QOC}}(t,\boldsymbol{\beta}) = H_0(t) + f(t,\boldsymbol{\beta})\mathcal{O}_{\text{opt}}  \;\;.
\end{equation}
The first term, usually addressed as the \emph{bare} Hamiltonian, describes the dynamics of the target system without regard for optimization. The second term is an additional driving term, which makes use of the operators  $\mathcal{O}_{\text{opt}}$ to provide additional degrees of freedom used by the optimizer subroutine.

\subsection{Counterdiabatic Driving}
\label{CD}

Various protocol designs, which are called \emph{shortcuts to adiabaticity}, have been proposed in order to mitigate the limitations prompted by the adiabatic theorems. The general target is to shorten as much as possible the evolution time and avoid gap closures in the spectrum.
Among the shortcuts-to-adiabaticity methods, Counterdiabatic Driving (CD) \cite{CD1,CD2,CD_Berry} is a promising candidate, which is formally able to overcome this problem through an ingenious choice of the additional driving term.
The basic idea behind CD is to boost any adiabatic process by adding a CD Hamiltonian (the \emph{adiabatic gauge potential}) that suppresses transitions between the system eigenstates:
\begin{equation*}
    H_{CD}(t) = H_0(t) + 
    \underbrace{i\hbar\sum_n
    \left( \ket{\partial_t n}\bra{n} - \bra{n}\ket{\partial_t n} \ket{n}\bra{n} \right)}_
    {\mathcal{A}_{\text{AGP}}(t)},
\end{equation*}
with $\ket{n}\equiv\ket{n(t)}$ being the $n$-th eigenstate of the instantaneous Hamiltonian $H_0(t)$.
To construct the additional driving term $\mathcal{A}_{\text{AGP}}$ it is necessary to have prior knowledge of all the eigenstates at all times during the system's dynamics.
This represents a huge limitation from the computational and even from the experimental point of view.

To overcome this complexity, it is common use to approximate the adiabatic gauge potential $\mathcal{A}_{\text{AGP}}$  using  suitable, local ansätze. The protocols designed are commonly referred as Local Counterdiabatic Driving (LCD) \cite{LCD_Dries,LCD_Gjonbalaj,LCD_Kolodrubetz}. 
The system dynamics is controlled by the time-dependent Hamiltonian
\begin{equation}
\label{eq:CDhamiltonian}
    H_{\text{CD}}(t) = H_0(t) + 
    \underbrace{\sum_j \alpha_j(t)\mathcal{O}_{\text{LCD}}^{(j)}
    }_{\mathcal{A}(t)}
\end{equation}
\begin{equation}
\label{eq:agpcondition}
    s.t. \qquad \mathcal{A}(t)\simeq\mathcal{A}_{\text{AGP}}(t)
\end{equation}
where $\{\mathcal{O}_{\text{LCD}}\}$ is a set of LCD operators. In this framework, $\boldsymbol{\alpha}(t)$ is optimized to approximate the adiabatic gauge potential \cite{variational_AGP_Campbell,variational_AGP_Saberi}, see Eq.~\ref{eq:agpcondition}.

\subsection{Counterdiabatic Optimized Local Driving}
\label{COLD}
Counterdiabatic Optimized Local Driving (COLD) is a method \cite{COLD} combining QOC and LCD. We
consider the CD Hamiltonian of Eq.~\ref{eq:CDhamiltonian} and replace the bare Hamiltonian $H_0$ with the QOC Hamiltonian $H_{\text{QOC}}$ of Eq.~\ref{eq:QOCHamiltonian}.
In the context of ASP and QA, the control function must vanish at boundary, i.e., $f(0,\boldsymbol{\beta}) = f(\tau,\boldsymbol{\beta}) = 0$.

Explicitly, the Hamiltonian is written as
\begin{equation}
    \mathcal{H}_{COLD}(t) = 
    \underbrace{\left( H_0(t) 
    + f(t,\boldsymbol{\beta})\mathcal{O}_{\text{opt}} 
    \right)}_{H_\beta(t)}
    +\,\boldsymbol{\alpha}(t, \boldsymbol{\beta})\mathcal{O}_{\text{LCD}}.
\end{equation}
Most importantly, we remark that $\boldsymbol{\alpha}(t) \rightarrow \boldsymbol{\alpha}(t,\boldsymbol{\beta})$, i.e. the optimization of the gauge approximation through $\boldsymbol{\alpha}(t)$ is now depending on the choice of the QOC parameters $\boldsymbol{\beta}$.

The coefficient of the QOC term is then given by the control function
\begin{equation*}
    f(t,\boldsymbol{\beta}) = \sum_{k=1}^{N_k}\beta^k\sin\left( \pi kt/\tau \right)
    \;\;
\end{equation*}
which represents a parameterized pulse. The optimization task consists in determining the coefficient $\beta^k \in \boldsymbol{\beta}$ of the \emph{k}th frequency of the control function.

In ASP and QA problems, it is common to write the bare Hamiltonian as
\begin{equation*}
    H_0(t) = H_i + \lambda(t)(H_f - H_i)
\end{equation*}
where $\lambda(t): [0,\tau] \rightarrow [0,1]$ is a monotonically increasing function (called \emph{schedule function}), $H_i\equiv H_0(0)$ and $H_f\equiv H_0(\tau)$ trivially.
It is common to use a linear schedule function $\lambda(t) = t/\tau$, but instead we focus on schedule function of the form
\begin{equation*}
    \lambda(t) = \sin^2\left(
    \frac{\pi}{2} \sin^2
        \left(\frac{\pi t}{2\tau}\right) 
    \right),
\end{equation*}
whose first $\dot\lambda$ and second derivative $\ddot\lambda$ vanish at the end\hyp points of the protocol.

Once that $H_\beta$ is determined (i..e. $\boldsymbol{\beta}$ is fixed), the parameters $\boldsymbol{\alpha}$ are optimized to realize the condition of Eq.~\ref{eq:agpcondition}.
For this reason, one could think of the adiabatic gauge potential ansatz being path\hyp dependent. The approximation of the adiabatic gauge potential is realized following the methods of Ref.~\cite{Sels_2017}. Briefly, it consists of defining the quantity
\begin{equation}
    \label{eq:Gdefinition}
    G = \partial_t H_\beta + \frac{i}{\hbar}[\mathcal{A}, H_\beta],
\end{equation}
which satisfies the closed-form equation $[G, H_\beta] = 0$.
Eventually, the condition of Eq.~\ref{eq:agpcondition} can be cast as the minimization of the Hilbert-Schmidt norm of $G$, or equivalently to the minimization of the action
\begin{equation}
    \label{eq:minaction}
    S(\mathcal{A}) = \Tr\left[G(\mathcal{A})^2\right]
    \;\;.
\end{equation}
This condition is equivalent to a set of $\text{dim}(\boldsymbol{\alpha})$ equations that have to be solved $\forall t \in [0,\tau]$.

\subsection{Extension for spin system applications}

It is clear that the minimization of Eq.~\ref{eq:minaction} has to be carried out specifically for each combination of system ($H_0$), control terms ($\mathcal{O}_{\text{opt}}$) and gauge potential ansatz ($\mathcal{A}$). For instance, the original work \cite{COLD} carries out the computations for a linear chain 1D Ising model. In general, this procedure is non-trivial, especially when facing more complex spin systems.

Our main contribution to the application of COLD consists in finding a suitable AGP for the ANNNI model, through a symbolical  framework for solving Eq.~\ref{eq:minaction}, which is implemented through an extension of the \emph{SymPy} library \cite{SymPy}.  The resulting framework \cite{colder} is able to automatize the workflow of COLD for arbitrary systems and ansatz choice.

Another technical enhancement introduced in this paper consists in the replacement of the Powell loss optimizer \cite{powell_efficient_1964} by a Bayesian optimizer (BO) \cite{brochu2010tutorial}. BO efficiently samples the loss function by changing the parameter $\boldsymbol{\beta}$ in a pre-defined interval, and returns the optimal parameter. While it is experienced that BO does not reduce drastically the number of iterations required to select an optimal parameter, we observe empirical advantage in escaping from local minima. In this regard, other optimization strategies might get trapped in local minima, and would required to run the optimization several times with different configuration. 

While the landscapes of the infidelity loss function, see Eq.~\ref{eq:loss_fidelity}, will naturally drives towards states with high fidelity, it still requires access to the exact state. Even if this is could lead to faster protocol by optimizing on a slow driving schedule, this is not optimal as the target state is usually not available. For this reason, we instead propose to minimize the energy loss function 
\begin{equation}
    \label{eq:loss_energy}
    \mathcal{C}_e(\boldsymbol{\beta}) \equiv \varepsilon = \frac{\bra{\psi(\tau, \boldsymbol{\beta})}H(\tau)\ket{\psi(\tau,\boldsymbol{\beta})} - E_{min}}{E_{max} - E_{min}}
\end{equation}
as often used in the context of the variational quantum eigensolver \cite{vqe_original,VQE_Gambetta,li6,LMG_grossi}. We observe that the energy loss function yields the same results as the infidelity one, while having the advantage of being computable without knowing the exact solution. One downsize is that it also gives less information about the success of the optimization, as no lower bounds are known a-priori.

Having introduced these tools, we report application of the COLD protocol on the ground-state preparation of systems exhibiting non-trivial phase diagrams, such as the ANNNI, the XXZ model and an all-to-all connected Haldane–Shastry model.

\section{Results}
\label{results}
Here, the application of the COLD method for the ground-state preparation of some spin systems is discussed. In general, the relative improvement of the prepared state fidelity (which is defined as $\mathcal{R}$) remarkably follows a pattern depending on the quantum phase of the prepared ground-state, with consistent enhancement in non-ferromagnetic phases.\\

As a reminder, in a typical QA protocol the spin system is prepared in the ground-state of $H_i = -\sum_i^N\sigma^x_i$ and the evolution is carried out towards a target Hamiltonian $H_f^{\text{(sys)}}$, which identifies the specific model (see Eqs. \ref{eq:modelANNNI}, \ref{eq:modelXXZ} and \ref{eq:modelHS}).
All the final Hamiltonians have two free parameters that are scanned in order to prepare ground-states in the different phases. The COLD optimization is therefore carried out independently for each parameter combination.

The prepared state energy is minimized by tweaking one coefficient of the control function (thus $N_k=1$). The total evolution time is set to be $\tau = 0.01$. A chain of $N=5$  spins is considered, as a first baseline for all benchmarks, before eventually considering larger spin systems.

\subsection{Ansatz for the AGP}
As previously mentioned, the control operators $\mathcal{O}_{opt}$ play a fundamental role, as well as the choice of a suitable ansatz $\mathcal{A}$ for the gauge potential. The control operators are selected following a manual search through $d$\hyp local ansätze. A simple local control operator $\sigma_z$ on each spin seems to be sufficient to reach high fidelity for the ANNNI model. However, this choice is not effective when applied to the XXZ and the HS models. Instead, better results are obtained when choosing non-local driving term, in particular next-nearest-neighbor $\sigma_z$ interactions. \\

Hence, the choice of the ansatz plays a crucial for the success of LCD and COLD. From a theoretical point of view, it is fundamental to assure that the commutator between the ansatz and the $H_\beta$ Hamiltonian in Eq.~\ref{eq:Gdefinition} does not vanish. 
Furthermore, experimental considerations are also an important factor. Indeed, local interactions are easier to implement and map into real hardware, and for this reason, it is preferable to keep the ansatz relatively simple to avoid introducing an expensive overhead. 

The simplest ansatz which does not commute with the original schedule Hamiltonian is a local field on the $y$ direction:
\begin{align}
    \mathcal{A}_{\text{local}} =&\; \alpha_1\sum_i\sigma_i^y
    \;\;.
\end{align}
Following the footsteps of Ref.~\cite{COLD}, a second order ansatz (with two\hyp body) is introduced:
\begin{align}
    \mathcal{A}_{\text{near}} =&\; \alpha_1\sum_i\sigma_i^y 
    + \alpha_2\sum_i\left( \sigma_i^x\sigma_{i+1}^y + \sigma_i^y\sigma_{i+1}^x \right) \nonumber\\
    &+ \alpha_3\sum_i\left( \sigma_i^y\sigma_{i+1}^z + \sigma_i^z\sigma_{i+1}^y \right),
\end{align}
with next-neighbor interaction, or with next-nearest-neighbor to increase the expressivity
\begin{align}
    \mathcal{A}_{\text{next}} =&\; \alpha_1\sum_i\sigma_i^y \nonumber + \alpha_2\sum_i\left( \sigma_i^x\sigma_{i+1}^y + \sigma_i^y\sigma_{i+1}^x \right) \nonumber\\
    &+ \alpha_3\sum_i\left( \sigma_i^y\sigma_{i+1}^z + \sigma_i^z\sigma_{i+1}^y \right) \nonumber\\
    &+ \alpha_4\sum_i\left( \sigma_i^x\sigma_{i+2}^y + \sigma_i^y\sigma_{i+2}^x \right) \nonumber\\
    &+ \alpha_5\sum_i\left( \sigma_i^y\sigma_{i+2}^z + \sigma_i^z\sigma_{i+2}^y \right).
\end{align}

From a theoretical standpoint, a more complex ansatz is expected to perform better, since the analytical gauge potential can be retrieved by considering an ansatz with all possible interactions, a point which is indeed observed in this study. Therefore, we shift our paradigm on simple local ansätze, that are still achieving reasonable fidelities. 

In order to quantify the improvement of COLD in the context of QA, we compute the fidelity and the normalized energy of the prepared state in different settings: using only QOC, only LCD, both or with unassisted annealing (UA), when no additional protocols are used.
We defined a success metric, $\mathcal{R}$, as the fidelities ratio between COLD and UA annealing schedules:
\begin{equation}
    \mathcal{R} \equiv \mathcal{F}_{COLD}/\mathcal{F}_{\text{UA}}
    \;\;.
\end{equation}
An high $\mathcal{R}$ is indicative of an higher relative improvement of fidelity with respect to trivial unassisted schedules. This indicator does not replace the absolute fidelities $\mathcal{F}_x$, but acts instead as a complementary metric, and we report both of them. Even if the fidelity is the commonly metric used to quantify the success of ASP protocols, and is a parameter of the QPE algorithm, we argue that fidelity can under estimate the quality of a state. Hence, the fidelity is expected to degrade drastically with the system size. On the other hand, the energy can be sufficient to understand the distance between the prepared and target state.

\subsection{Results on the ANNNI model}

\begin{table*}[t]
\begin{minipage}{0.5\textwidth}
    \includegraphics[width=\textwidth]{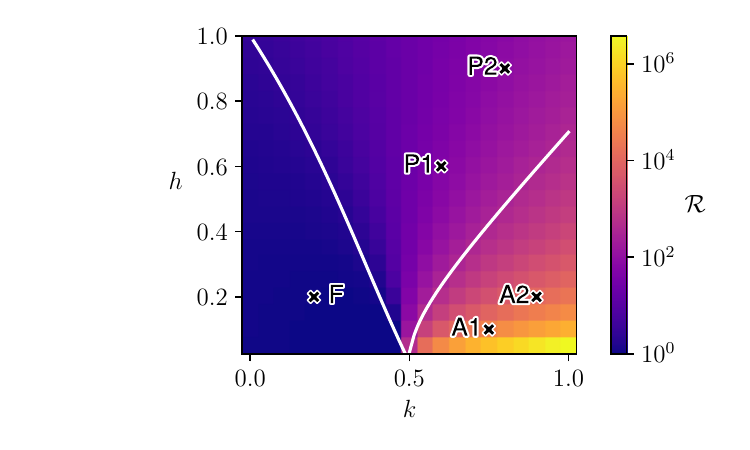}
\end{minipage}\hfill
\begin{minipage}{0.5\linewidth}
    \centering
    
    \begin{tabular}{||c|c|c||c|c|c}
    label& $k$ & $h$ & $\mathcal{F}_{\text{UA}}$ & $\mathcal{F}_{LCD}$ & $\mathcal{F}_{COLD}$\\\hline
F& 0.20	&0.2	&$0.4897$	&$0.5001$		&$0.5230$\\
P1& 0.60	&0.6	&$1.7671\cdot10^{-2}$	&$0.2701$	&$0.9386$\\
P2& 0.80	&0.9	&$1.0881\cdot10^{-2}$	&$0.5031$   &$0.9609$\\
A1& 0.75	&0.1	&$2.0579\cdot10^{-5}$	&$3.8196\cdot10^{-4}$	&$0.7778$\\
A2& 0.90	&0.2	&$7.9326\cdot10^{-5}$ &$4.2983\cdot10^{-3}$	&$0.8525$\\
    \end{tabular}
\end{minipage}
\caption{\label{tab:annni_samples}ANNNI model, ground-state preparation for $N=5$ spins. Several configurations $(k,h)$ are sampled from the phase diagram on the left, and the absolute fidelities are reported in the table.}
\end{table*}

\begin{table*}[t]
    \centering
    \begin{tabular}{cr|c|c|c|c|c}
        method&metric&F&P1&P2&A1&A2\\ \hline
        \multirow{2}{*}{UA} & $\varepsilon$ &   0.5 &    0.5 &    0.5 &      0.5 &      0.5  \\
         & $\mathcal{F}$  &  0.49 & 0.0177 & 0.0109 & 2.06$\cdot10^{-5}$ & 7.93$\cdot10^{-5}$\\ \hline
        \multirow{2}{*}{COLD + $\mathcal{A}_{\text{local}}$} & $\varepsilon$
            & 0.141 &  0.141 &  0.141 &    0.141 &    0.141 \\
        & $\mathcal{F}$ & 0.121 &  0.689 &  0.721 &    0.363 & 0.4 \\ \hline
        \multirow{2}{*}{COLD + $\mathcal{A}_{\text{near}}$} & $\varepsilon$
            & 0.174 &  0.111 & 0.0938 &    0.164 &    0.142  \\
        & $\mathcal{F}$ & 0.523 &  0.782 &  0.785 &    0.455 &    0.485 \\ \hline
        \multirow{2}{*}{COLD + $\mathcal{A}_{\text{next}}$} & $\varepsilon$
            & 0.148 & 0.0467 &  0.023 &   0.0721 &   0.0435  \\
        & $\mathcal{F}$ & 0.436 &  0.938 &  0.961 &    0.775 &     0.85  \\ \hline
    \end{tabular}
    \caption{ANNNI model simulations carried out using all the three ansatz choices. We report the normalized prepared-state energy $\varepsilon$ and its fidelity $\mathcal{F}$.}
    \label{tab:annni-comparison-ansatz}
\end{table*}

\begin{figure}[h]
    \begin{subfigure}[t]{0.5\textwidth}
    \includegraphics[width=\textwidth]{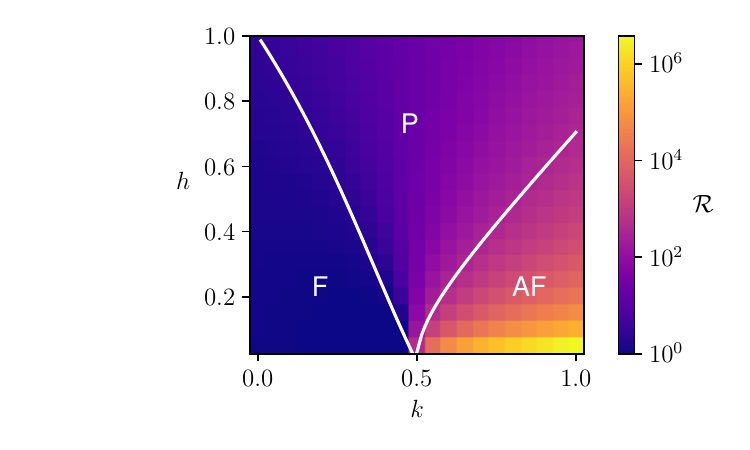}
    \caption{\label{fig:annni_grid}Relative improvement of fidelity using COLD. 
    }
    \end{subfigure}

    \begin{subfigure}[t]{0.5\textwidth}
    \includegraphics[width=\textwidth]{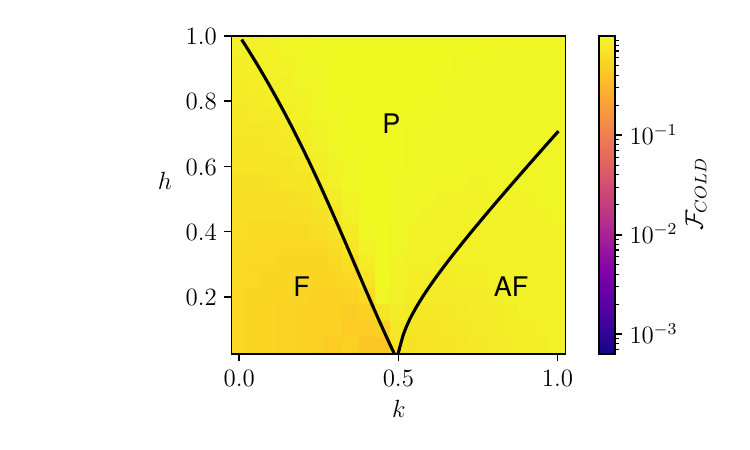}
    \caption{\label{fig:annni_abs} Absolute fidelity achieved by COLD.}
    \end{subfigure}
    
\caption{\label{fig:annni_cold}Results of ground-state preparation of an ANNNI model in a QA setup. The Hamiltonian of Eq.~\ref{eq:modelANNNI} is prepared for different values of $h$ and $k$. Simulations for $N=5$ spins, using the ansatz $\mathcal{A}_{next}$.}
\end{figure}

\begin{table*}[t]
\begin{minipage}{0.33\textwidth}
    \hspace*{5mm}\includegraphics[width=1.1\textwidth]{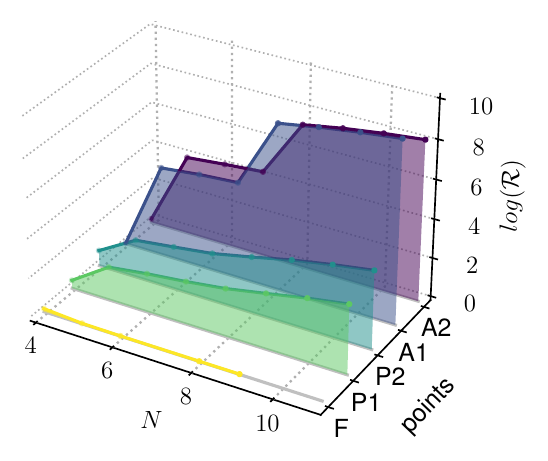}
\end{minipage}\hfill
\begin{minipage}{0.65\linewidth}
    \centering
    \begin{tabular}{c|c|c|c|c|c}
    \multirow{2}{*}{label}& \multicolumn{5}{c}{$\mathcal{R}$}\\
    &$N=4$	&$N=5$	&$N=7$	&$N=9$	&$N=11$\\\hline
    F	&$1.437$	&$1.067$	&$1.000$	&$1.037$	&$1.000$\\
    P1	&$2.647$	&$53.12$	&$1.611\cdot10^{2}$	&$6.979\cdot10^{2}$	&$3.646\cdot10^{3}$\\
    P2	&$1.089$	&$3.780\cdot10^{4}$	&$8.700\cdot10^{4}$	&$6.319\cdot10^{8}$	&$2.119\cdot10^{9}$\\
    A1	&$6.350$	&$88.30$	&$2.827\cdot10^{2}$	&$2.124\cdot10^{3}$	&$1.044\cdot10^{4}$\\
    A2  &$1.510$    &$1.075\cdot10^{4}$	&$2.485\cdot10^{4}$	&$4.94\cdot10^{7}$	&$1.619\cdot10^{8}$\\
    \end{tabular}
\end{minipage}
\caption{\label{tab:annni_N}ANNNI model, $\mathcal{F}_{COLD}/\mathcal{F}_{\text{UA}}$ for various $N$.}
\end{table*}

The ANNNI model is an extension of the Ising model which introduces next-nearest-neighbor couplings between spins. The target Hamiltonian is written as
\begin{equation}
    \label{eq:modelANNNI}
    H_f^{\text{(ANNNI)}} = -J^*\sum_i \sigma^x_i \sigma^x_{i+1} + k\sum_i \sigma^x_i \sigma^x_{i+2} + h\sum_i \sigma^z_i
    \;\;.
\end{equation}
The coupling constant $J^*$ (which is set to 1) sets the energy scale, while the dimensionless parameters $k$ and $h \in [0,1]$ account for the next-nearest-neighbor interaction and the transverse magnetic field, respectively. 
The presence of opposite signs in the nearest and next-nearest interactions, resulting in either ferromagnetic or antiferromagnetic exchanges within the system, gives rise to magnetic frustration. Consequently, the ANNNI model provides an avenue for exploring the interplay between quantum fluctuations induced by the transverse magnetic field and frustration. The phase diagram exhibits significant intricacy, with the confirmation of three distinct phases, ferromagnetic (F), paramagnetic (P) and anti-ferromagnetic (AF), delineated by two second-order phase transitions \cite{PRB_ANNNI,PRB_DMRG_ANNNI,ANNNI_report, Monaco_ANNNI}.

The set of actively controlled operators is given by a local field homogeneously applied to the spin chain. \begin{equation*}
    \mathcal{O}_{opt}^{(\text{ANNNI})} = \sum_i \sigma^z_i
\end{equation*}

\subsubsection{Effect of the optimization}
In the following, we report the difference achieved by QOC, LCD, COLD and the UA annealing schedules.

Figure ~\ref{fig:annni_grid} shows the success metric $\mathcal{R}$ for the ground-state preparation of the ANNNI model through different combinations of parameters ($k$,$h$) using the local APG ansatz.
Remarkably, the greatest improvement is obtained in the preparation of the paramagnetic and anti-ferromagnetic phases, improving by 3 to 6 orders of magnitude.
From looking at Figure \ref{fig:annni_abs}, it emerges that the absolute fidelity obtained using COLD $\mathcal{F}_{COLD}$ is roughly homogeneous across all the phases of this model, with values greater than $5\cdot 10^{-1}$.\\

A deeper analysis is performed on a selection of points across the phase diagram, and the results are reported in Table~\ref{tab:annni_samples}.
The data hints that the ground-state in the ferromagnetic phase is easily reachable, as the unassisted schedule already reaches a fidelity of $0.5$ which does not increment significantly by using either LCD or COLD.
Instead, a noticeable difference emerges in the paramagnetic phase, where unassisted schedules present small absolute fidelities, which are slightly improved by LCD and are eventually maximized to $\sim 0.95$ by COLD.
As of the anti-ferromagnetic phase, the unassisted annealing schedules exhibit almost vanishing fidelities, which are boosted by COLD to $\mathcal{F}_{COLD} \simeq 0.85$.
Since the fidelity of LCD is equal to the bare annealing fidelity, we conclude that QOC is crucial to prepare anti-ferromagnetic ground-states in which the final transverse field is null ($h=0$).\\


\subsubsection{Effect of the ansatz}

In the following, we study how the improvement is affected by the choice of the AGP ansatz. We consider the same selection of points in the parameter space of the ANNNI model and execute the simulation for the three AGP ansätze introduced in the previous section. The results are shown in Table \ref{tab:annni-comparison-ansatz}.
While it is true that the prepared state fidelities over the paramagnetic and antiferromagnetic phases increase when amore complex ansatz is used, the improvement over the UA protocol remains dominant, and the fidelities obtained with a more complex approach are within the same order of magnitude. Thus, the simplest ansatz remains a solid option for the ANNNI model.
From the energy point of view, the indicator $\varepsilon$ validates the COLD protocol in any setup. Even though the fidelities lie in the same order of magnitude, a more complex ansatz leads to sensibly lower energies.\\
\subsubsection{Effect of the system size}

In this section, we study the effect  how the improvement how the fidelity scales with the system size $N$.
Table \ref{tab:annni_N} reports the success metric $\mathcal{R}$ for $5\leq N\leq 11$.
We remark that the improvement is generally increasing with the system size. This is due to absolute fidelities of the unassisted schedules which decrease in larger system, while the COLD fidelities, remains fairly constant.\\


\subsubsection{Schedules inheritance}
A final result is worth to be mentioned. In general, one could think that the COLD optimizations over the same system (and parameter settings) have to be carried out independently for different values of $N$.
However, a pattern in the optimized parameters hints some form of \emph{inheritance} of the optimized parameters towards systems of larger size.

This observation would be of critical interest in the optimization of larger systems, since the cost of each QA simulation can be computationally expensive. Indeed, it would be possible to inherit the $\boldsymbol{\beta}$ parameters from optimizations carried out on small values of $N$, eventually using such parameter on larger systems without executing the optimization routine from scratch.

This pattern has shown to be beneficial in the case of the ANNNI model. However, no guarantees can be given in general.
Indeed, further tests on XXZ an HS systems have proved that this inheritance mechanism is not always as effective as in the ANNNI model. 

To give a successful example of inheritance, Table \ref{tab:annni-inheritance} shows the optimized parameter for the ANNNI model in the $A1$ configuration. The fidelities obtained with independent optimizations only present a small improvement of the order $10^{-3}$ over the inherited fidelities.
Furthermore, the independently optimized parameters are close to the inherited parameter, hinting that the independent optimizations are a second order correction to the inherited parameter.

\begin{table}[h]
\centering
\begin{tabular}{c|c|c|c|c}
& \multicolumn{2}{c|}{inherited from $N=5$} & \multicolumn{2}{c}{independent optimization}\\
$N$  & $\beta$ & $\mathcal{F}_{COLD}$  & $\beta$   & $\mathcal{F}_{COLD}$ \\\hline
6  & \multirow{4}{*}{3.8607} & 0.71919 & 3.3080 & 0.71932 \\
7  &                         & 0.64658 & 3.1832 & 0.64675 \\
9  &                         & 0.26225 & 4.2011 & 0.26228 \\
11 &                         & 0.33197 & 3.5228 & 0.33201
\end{tabular}
\caption{\label{tab:annni-inheritance}Test of optimized parameter inheritance on the $A1$ configuration of the ANNNI model.}
\end{table}

\subsection{Further tests on spin systems}

\begin{table*}
\begin{minipage}{0.37\textwidth}
    \hspace{-14mm}
    \includegraphics[width=1.20\textwidth]{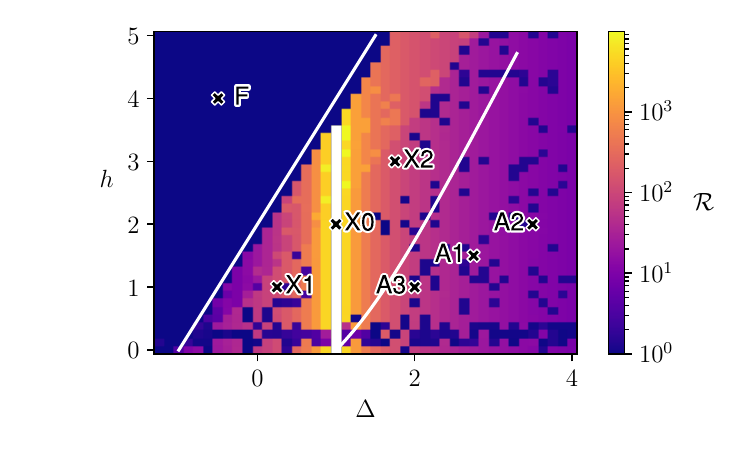}
\end{minipage}\hfill
\begin{minipage}{0.62\linewidth}
    \centering
    \begin{tabular}{||c|c|c||c|c|c|c}
     & & & $\mathcal{F}_{\text{UA}}$ & $\mathcal{F}_{LCD}$ & $\mathcal{F}_{COLD}$ & $\mathcal{F}_{COLD}$\\
    label & $\Delta$ & $h$ & - & with $ \mathcal{A}_{local}$&$\mathcal{A}_{local}$& $\mathcal{A}_{next}$\\\hline
    F& -0.50	&4.0	&$3.13\cdot10^{-2}$	&$0.972$	&$0.9730$ & $0.9995$\\
    X0& 1.00	&2.0	&$8.21\cdot10^{-32}$	&$4.95\cdot10^{-32}$	&$0.0837$ & $0.1160$\\
    X1& 0.25	&1.0	&$7.31\cdot10^{-4}$	&$4.38\cdot10^{-5}$	&$0.0886$ & $0.1485$\\
    X2& 1.75	&3.0	&$5.92\cdot10^{-4}$	&$5.92\cdot10^{-4}$	&$0.0767$ & $0.0851$\\
    A1& 2.75	&1.5	&$2.70\cdot10^{-3}$	&$2.70\cdot10^{-3}$	&$0.0664$ & $0.1570$\\
    A2& 3.50	&2.0	&$4.69\cdot10^{-3}$	&$4.69\cdot10^{-3}$	&$0.0601$ & $0.1651$\\
    A3& 2.00	&1.0	&$1.01\cdot10^{-3}$	&$1.01\cdot10^{-3}$	&$0.0739$ & $0.0699$
\end{tabular}

\end{minipage}
\caption{\label{tab:xxz_samples}XXZ model, ground-state preparation for $N=5$ spins. Several simulations $(\Delta,h)$ are sampled from the phase diagram on the left, and the absolute fidelities are reported in the Table on the right. Points for $\Delta=1$ and $h<4$ are intentionally left blank as the relative improvement $\mathcal{R}$ is greater than $10^{29}$.}
\end{table*}

\begin{table*}
\begin{minipage}{0.35\textwidth}
    \hspace{-2.0cm}
    \includegraphics[width=1.3\textwidth]{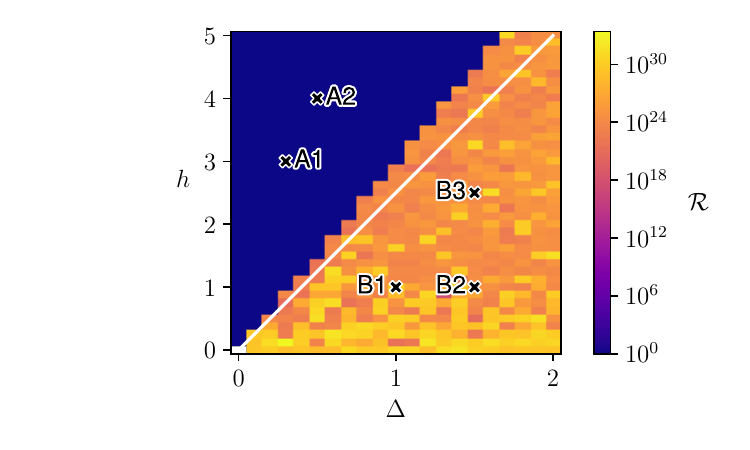}
\end{minipage}\hfill
\begin{minipage}{0.63\linewidth}
\centering
\begin{tabular}{||c|c|c||c|c|c|c}
     &  &  & $\mathcal{F}_{\text{UA}}$ & $\mathcal{F}_{LCD}$ & $\mathcal{F}_{COLD}$&$\mathcal{F}_{COLD}$\\	
    label & $\Delta$ & $h$ & - & with $\mathcal{A}_{local}$ & $\mathcal{A}_{local}$&$\mathcal{A}_{next}$\\\hline		
    A1& 0.3&	3.0&	$3.13\cdot10^{-2}$	&$1.00$	&$1.00$&$1.00$\\
    A2& 0.5&	4.0&	$3.13\cdot10^{-2}$	&$1.00$	&$1.00$&$1.00$\\
    B1& 1.0&	1.0&	$2.45\cdot10^{-32}$	&$5.07\cdot10^{-32}$	&$0.0287$&$0.0459$\\
    B2& 1.5&	1.0&	$1.93\cdot10^{-33}$	&$1.35\cdot10^{-31}$	&$1.27\cdot10^{-8}$&$0.0159$\\
    B3& 1.5&	2.5&	$8.60\cdot10^{-32}$	&$3.15\cdot10^{-32}$	&$4.27\cdot10^{-7}$&$0.0755$
\end{tabular}
\end{minipage}
\caption{\label{tab:hs_samples}HS model, ground-state preparation for $N=5$ spins. Several simulations $(\Delta,h)$ are sampled from the phase diagram on the left, and the absolute fidelities are reported in the Table on the right.}
\end{table*}

In this section, COLD is applied to more complex spin systems, such as the XXZ and the Haldane-Shastry models.

The main difference between the simulations on the ANNNI model, is a new choice for the QOC operators. As mentioned earlier, preliminary tests hinted that COLD is more effective on these models when two-body interactions are included. Hence, a simple control over local fields has not shown to be effective as it was in the ANNNI model.
Thus, we have settled to control next-nearest-neighbor couplings along the quantization axis $\vec{z}$ pointing in the field direction:
\begin{equation*}
    \mathcal{O}_{opt}^{(\text{XXZ})} =\mathcal{O}_{opt}^{(\text{HS})}= \sum_i \sigma^z_i\sigma^z_{i+2}
\end{equation*}

\subsubsection{XXZ model}

The XXY spin model presents two-body couplings along the field direction $\vec{z}$ with coefficient $\Delta$, as well as in the orthogonal plane (controlled by the fixed coefficient $J^*=1$). The variation of the free parameters $\Delta$ and $h$ allows the formation of ferromagnetic (F), superfluid (XY) and anti-ferromagnetic phases (AF).

\begin{align}
    \label{eq:modelXXZ}
    H_f^{\text{(XXZ)}} =\;& J^*\left( \sum_i \sigma^x_i \sigma^x_{i+1} + \sum_i \sigma^y_i \sigma^y_{i+1} \right) \nonumber\\
    &+ \Delta\sum_i \sigma^z_i \sigma^z_{i+1} + h\sum_i\sigma^z_i
\end{align}

The results are structured as in the previous section, where some points in the different phases have been selected for deeper analysis,  and are reported in Table \ref{tab:xxz_samples}.

The results of the UA schedule $\mathcal{F}_{\text{UA}}$ exhibit the same trend of the ANNNI model, hinting that non-ferromagnetic phases are not well prepared with the bare annealing protocol.
The ferromagnetic point $F$ is boosted up to a fidelity of $0.999$ with the contribution of LCD. However, it seems that LCD alone is not sufficient to increase the fidelity in the preparation of the other phases. Instead, it is COLD which assumes a dominant role in those regions of the phase diagram, leading to the preparation of states with fidelity $\mathcal{F}_{COLD}\simeq 10^{-1}$ when the ansatz $\mathcal{A}_{next}$ is used.
The simpler ansatz $\mathcal{A}_{local}$ is still effective, but the absolute fidelities are instead in the order of $10^{-2}$.

A remarkable improvement is achieved in the configuration $X0$, as well the other points in the region $\Delta=1$ and $h<4$.
A quick look at the figure attached to Table \ref{tab:xxz_samples} shows an increasingly higher value of the success metric $\mathcal{R}$ when approaching  $\Delta=1$ in the superfluid phase.

From a statistical mechanics point of view, the line $\Delta=1$ is significant, as the symmetry of 
the coupling in the field direction becomes homogeneous to the couplings in the $xy-$plane transverse to the magnetic field. Thus, the model presents a rotational SU(2) symmetry in the spin couplings ,eventually broken by the magnetic term. Therefore, the system is an isotropic ferromagnet with a gapless spectrum.

This specific case has been motivating in the choice of the third and last model subject of our studies, which will present the same symmetry.

\subsubsection{Haldane-Shastry model}

%

The Haldane-Shastry (HS) model describes a spin chain with long-range antiferromagnetic interactions. This model is exactly solvable using the asymptotic Bethe ansatz and features a spin-liquid ground-state \cite{Morong_2023}.

\begin{align}
    \label{eq:modelHS}
    H_f^{\text{(HS)}} =\; & \Delta \sum_i\sum_{j>i} \frac{1}{|r_i-r_j|^2} \left( \sigma^x_i \sigma^x_{j} +\sigma^y_i \sigma^y_{j} + \sigma^z_i \sigma^z_{j}\right)\nonumber\\
    &+ h\sum_i\sigma^z_i
\end{align}
The spins are supposed to be equally spaced in a unitary circle. The coefficient of the all-to-all interactions are determined by the inverse square of the site distance. 
For this reason, we can set $r_k = e^{i2\pi k/N}$.
As in the case of the XXZ model, the most effective control term choice has been selected in the exploratory phase of this work, eventually settling to control over the next-nearest-neighbor interactions.

The results of the COLD application to this model hints, once again, that the final ground-state fidelity and the relative improvement $\mathcal{R}$ is correlated to the phase of the prepared state.

Using the data of Table~\ref{tab:hs_samples}, two trends can be distinguished.
The points $A1$ and $A2$, belonging to the ferromagnetic phase, are subject to a small relative improvement. However, their absolute fidelity is saturated to $1$ when the preparation is carried out with LCD. Indeed, by looking at the absolute fidelities we conclude that LCD alone is responsible for this achievement, and COLD is not necessary to the preparation of the ferromagnetic states.

Instead, the points $B1$, $B2$ and $B3$ show an extremely high improvement ratio $\mathcal{R}\simeq 10^{31}$. LCD is not effective in the preparation of such states, as the absolute fidelity does not show any improvement. Nevertheless, COLD is crucial, pushing the fidelities up by many orders of magnitude, settling to $\mathcal{F}_{COLD}\simeq 10^{-2}$ when the ansatz $\mathcal{A}_{next}$ is used.\\

\section{Conclusions}
Achieving a high level of confidence in state preparation is of paramount importance in the era of quantum simulation. Especially, quantum critical systems pose an open challenge to implementation schemes relying their validity on adiabatic preparation. 
The difficulties arising in the control when implementing adiabatic computation is manifest when the system could be driven along phase transitions across a quantum critical point that can result into tunneling in excited levels. 

In this paper, we thoroughly discuss the application of extensions and benchmarking of counterdiabatic driving protocols (COLD) across various non-trivially integrable models that exhibit  rich phase diagrams. Our findings indicate that COLD consistently outperforms standard annealing methods, often by orders of magnitude, and achieves nearly perfect fidelity when applied to ferromagnetic states. Notably, COLD demonstrates good efficacy in addressing the ANNNI model, characterized by its non-integrable nature and diverse phase diagram. Furthermore, we show that optimized parameters from smaller models can be extended to higher-dimensional systems, emphasizing the pivotal role of optimized paths and adiabatic compensation in practical adiabatic state preparation applications. We introduce several enhancements, including the utilization of Bayesian optimization and higher-order approximate adiabatic gauge potentials, supported by a dedicated code package \cite{colder} for convenient evaluation.

Despite the success of COLD in the ANNNI model, its performance is less impressive when applied to the XXZ and HS chain models. While it still surpasses standard annealing protocols, the achieved fidelities fall short of the requirements for ASP in these contexts. Consequently, we advocate further research efforts aimed at identifying effective adiabatic gauge potentials for various models. 

\section*{Acknowledgments} FPB, OK, SV and MG are supported by CERN through the CERN Quantum Technology Initiative. AM is supported by Foundation for Polish Science (FNP), IRAP project ICTQT, contract no. 2018/MAB/5, co\hyp financed by EU Smart Growth Operational Program.

\bibliography{refs,bibliography_li6} 

\end{document}